\newcommand{\stareqn}{\renewcommand{\theequation}{\mbox{$*\arabic{equation}$}}}
\newcommand{\resteqn}{\renewcommand{\theequation}{\arabic{equation}}}

\newcommand{\One}{{\bf 1}}
\newcommand{\ag}{\alpha}
\newcommand{\bg}{\beta}
\newcommand{\cg}{\gamma}
\newcommand{\dg}{\delta}
\newcommand{\sg}{\sigma}

\newcommand{\di}{\partial}
\newcommand{\be}{\begin{equation}}
\newcommand{\ee}{\end{equation}}
\newcommand{\bearr}{\begin{eqnarray}}
\newcommand{\eearr}{\end{eqnarray}}
\newcommand{\Sg}{\Sigma}

\newcommand{\m}{\mu}
\newcommand{\n}{\nu}
\newcommand{\eg}{\epsilon}

\newcommand{\Dg}{\Delta}

\newcommand{\QED}{\rule{1mm}{3mm}}

\newcommand{\calP}{{\cal P}}

\documentstyle[12pt,psfig]{article}

\begin{document}

\title{A left-handed simplicial action for euclidean general relativity}
\author{Michael P. Reisenberger\\
	Facultad de Ciencias, Universidad de la Rep\'ublica\\
	Trist\'an Narvaja 1674, 11200 Montevideo, Uruguay\\
	and\\
	Erwin Schr\"odinger International\\
	Institute for Mathematical Physics\\
	Boltzmanngasse 9, A-1090, Wien, Austria\\
	and\\
	Center for Gravitational Physics and Geometry\\
	The Pennsylvania State University\\
	University Park, PA 16802, USA}
\maketitle

\begin{abstract}
An action for simplicial euclidean general relativity involving only 
left-handed fields is presented. The simplicial theory is shown to converge to
continuum general relativity in the Plebanski formulation as the simplicial 
complex is refined. This contrasts with the Regge model for which M. Miller
and Brewin have shown that the full field equations are much more restrictive
than Einstein's in the continuum limit. The action and field equations of the
proposed model are also significantly simpler then those of the Regge model 
when written directly in terms of their fundamental variables.

An entirely analogous hypercubic lattice theory, which approximates
Plebanski's form of general relativity is also presented.
\end{abstract}

\section{Introduction} 

It has been known for some time that in general relativity (GR) the 
gravitational field can be represented entirely by left-handed fields, i.e.
connections and tensors that transform only under the left-handed, or
self-dual subgroup of the frame rotation group.\footnote{In euclidean
GR the frame rotation group is $SO(4)$ which can be written as the tensor
product $SU(2)_R\otimes SU(2)_L$ (in terms of fundamental representations). Left 
handed tensors 
transform only under the $SU(2)_L$ factor. Examples are left handed 
spinors and self-dual antisymmetric tensors, i.e. tensors $a$ that satisfy
$a^{IJ} = \eg^{IJ}{}_{KL} a^{KL}$.}

The present paper presents a simplicial model of GR with an internal 
$SU(2)$ gauge symmetry which, at least in the continuum limit, corresponds
to the left handed frame rotation group. The gravitational field is 
represented by spin 2 and spin 1 $SU(2)$ tensors, and $SU(2)$ parallel 
propagators, associated with the 4-simplices, and with 2-cells and edges 
constructed from the 4-simplices, respectively.

This is meant to provide a step in the construction of a covariant path 
integral, or sum over histories, formulation of loop quantized GR. 
In Ashtekar's reformulation of classical canonical GR \cite{Ashtekar86},
\cite{Ashtekar87} the canonical variables are the left handed part of the spin 
connection on space and, conjugate to it, the densitized dreibein. The 
connection can thus be taken as the configuration variables, opening the door
to a loop quantization of GR \cite{Gambini86}, \cite{RSloops}, 
\cite{RSloops90}, \cite{ALMMT95}. In loop quantization one supposes that 
the state can be written
as a power series in the spatial Wilson loops of the connection (which
coordinatize the connections up to gauge), so the fundamental excitations
are loops created by the Wilson loop operators.
The kinematics of loop quantized canonical GR requires that geometrical
observables measuring lengths \cite{Thiemann_length}, areas \cite{discarea2}
\cite{AL96a}, and volumes \cite{discarea2}, \cite{AL96b}, \cite{L_volume},
\cite{Thiemann_volume} have discrete
spectra and finite, Plank scale lowest non-zero eigenvalues, suggesting
that GR thus quantized has a natural UV cutoff. One would therefore expect
that a path integral formulation of this theory would have, in addition
to manifest covariance, also reasonable UV behaviour.

A step toward such a path integral formulation is the construction of the
analogous formulation in a simplicial approximation to GR. Loop quantization
can be applied to any $SU(2)$ spacetime lattice\footnote{The lattice need not 
be hypercubic or regular in any way.} gauge theory in which the boundary is
a finite lattice and the boundary data is the connection on that lattice,
because in this case the states can always be expressed as power series
in Wilson loops.
Moreover, for local theories of this type it has been shown 
\cite{Reisenberger94} that 
the evolution operator can be written as a sum over the worldsheets of the 
loop excitations,\footnote{%
A lattice theory is ``local'' if the action of a region is the sum of the 
actions of basic cells (smallest subdivisions for which the action is defined) 
that make up the region, and the action of each basic cell depends only on 
fields living in the cell and on boundary data.

In \cite{Reisenberger94} the sum over worldsheets formulation is obtained 
for theories whose actions are functions of the connection only. However, 
if the connection is the only boundary datum for each basic cell, a local
action in terms of the connection only can be obtained by integrating out all
other variables in each cell.}
that is,
as a path integral. All that is needed for a path integral formulation of
loop quantized GR is a local lattice action for GR in terms of the left handed
part of the spin connection, and other fields, such that the connection is
the boundary data.

Plebanski \cite{Plebanski77} found precisely such an action in the continuum.
\footnote{The left handed part of the spin connection is also the boundary
data of the GR action of Samuel \cite{Samuel87}, and Jacobson and Smolin 
\cite{Jacobson_Smolin87}, \cite{Jacobson_Smolin88a}. However I shall not try 
to build a lattice
analogue of that action here.}
Here a simplicial lattice analogue of this action is presented. 
The corresponding path integral formulation of loop quantized simplicial
GR will appear in a forthcoming paper.

The present work might also be useful in numerical relativity.
Unlike the Regge model the present simplicial model has field equations
which reproduce those of general relativity in the continuum limit.
M. Miller \cite{Miller95} and Brewin \cite{Brewin95} have found linear
combinations of the Regge equations which {\em do} reproduce the
Einstein equations in the limit, however these do not define the extremum
of any known action, which presents a serious obstacle to any simplicial
path integral quantization based on the Regge model.  
A second advantage of the present model in numerical work that its
field equations are relatively simple in terms of the fundametal variables,
whereas the Regge equations, when written out in terms of edge lengths, are 
extremely complicated.
Finally, a simple modification of the new simplicial lattice action (given
at the end of the present paper) yields
a hypercubic lattice action for GR, which might lead to a simpler and faster
computer implementation than a simplicial model.

In section \ref{model} the simplicial model is presented. Field equations
and boundary terms are discussed in section \ref{fieldeq}. The continuum
limit is analysed in section \ref{continuum}. The results of M. Miller and Brewin
on the continuum limit of the Regge model are reviewed, and why it is
that the present simplicial model leads to field equations approximating those
of continuum GR while the Regge model does not is discussed.

Finally section \ref{comments}
contains some comments on the results, and also states the hypercubic action.
Appendix \ref{geometry} gives a metrical interpretation of the simplicial
field $e_s$, and in appendix \ref{lemmas} some lemmas used to establish
the continuum limit are proved.

\section{The model}\label{model}

Plebanski gave the following action for general relativity
(GR) in terms of left-handed fields \cite{Plebanski77}\cite{CDJM}:\footnote{
The definition of exterior multiplication used here is
$[a \wedge b]_{\ag_1 ... \ag_m\bg_1 ... \bg_n} = a_{[\ag_1 ... \ag_m}
b_{\bg_1 ... \bg_n]}$, where spacetime indices are labeled by lower
case greek letters $\{\ag,\bg,\cg,...\}$. Forms are integrated according 
to $\int_A a
= \int_A \eg^{u_1 ... u_m} a_{u_1 ... u_m}\ d^m\sg$ where $A$ is 
an $m$ dimensional manifold, $\sg^u$ are coordinates on $A$, the indices
$u_i$ run from $1$ to $m$, and $\eg^{u_1 ... u_m}$ is the $m$ dimensional
Levi-Civita symbol ($\eg^{12...m} = 1$ and $\eg$ is totally antisymmetric).}
\begin{equation}	\label{Plebanski_action}
I_P = \int \Sg_i \wedge F^i - \frac{1}{2} \phi^{ij} \Sg_i \wedge \Sg_j
\end{equation}
(The euclidean theory is obtained when all fields are real).
This action has internal gauge group $SU(2)$, with $\Sg$ a 2-form and an 
$SU(2)$ vector (spin 1), $F$ the curvature of an $SU(2)$
connection $A$, and $\phi$ a spacetime scalar and spin 2 $SU(2)$ tensor. 
The action is written in terms of the components of these fields in the
adjoint, or $SO(3)$, representation of $SU(2)$. (Indices in this representation
run over $\{1,2,3\}$ and will be indicated by lowercase roman letters
$\{i,j,k,l,...\}$). $\phi$ is thus represented by a traceless symmetric matrix
$\phi^{ij}$. 

On non-degenerate ($\Sg_k\wedge\Sg^k 
\neq 0$) solutions these fields can be expressed in 
terms of more conventional variables. $\Sg$ is the self-dual part of the
vierbein wedged with itself:\footnote{
Note that upstairs and downstairs $SO(3)$ indices are the same.}
\begin{equation}	\label{Sigma_e_wedge_e}
\Sg_i = 2[e \wedge e]^{+\,0i} \equiv e^0 \wedge e^i + \frac{1}{2}
	\eg_{ijk} e^j \wedge e^k,
\end{equation}
which transforms as a spin 1 vector under $SU(2)_L$, the left-handed 
subgroup of
the frame rotation group $SO(4) = SU(2)_R \otimes SU(2)_L$, 
and as a scalar under $SU(2)_R$. $A$ is the 
self-dual ($SU(2)_L$) part of the spin connection, and $\phi$ turns out to be
the left-handed Weyl curvature spinor.
The non-degenerate solutions correspond in this way exactly to the set of
solutions to Einstein's equations with non-degenerate spacetime metric.

Ashtekar's canonical variables are just the purely spatial parts of
$A$ and $\Sg$ (the dual of the spatial part of $\Sg$ is the densitized triad),
and, in the non-degenerate sector, the canonical  
theory derived from (\ref{Plebanski_action}) is identical to Ashtekar's
\cite{CDJM}\cite{Reisenberger95}. Since
this is precisely the sector of non-degenerate spatial metric it is of course
also equivalent to the ADM
theory \cite{ADM62}. However, when the metric is degenerate the canonical 
theory differs from Ashtekar's \cite{Reisenberger95}.
Since Plebanski's theory defines an extension of GR to degenerate geometries,
and this extension is not the only one possible, I will refer to this theory 
as Plebanski's theory.

\begin{figure}
\centerline{\psfig{figure=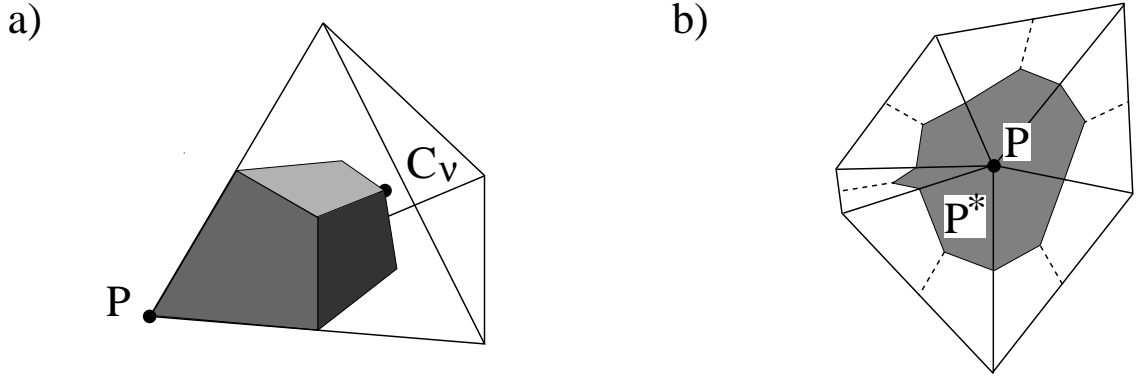,height=5cm}}
\caption[xxx]{Panel a) shows the corner cell $a_P$ associated with the vertex 
$P$ of a 3-simplex. Notice that the intersection of $a_P$
with any of the triangular faces incident on $P$ is itself the two dimensional
corner cell of $P$ in the triangular face, and that the six corners of 
$a_P$ are located at the centers of the simplices incident on $P$, defining, 
topologically, a cube.
\newline\newline
Panel b) shows the cell $P^*$ dual to the vertex $P$ in a two dimensional 
simplicial complex. The boundaries of other dual cells are indicated by 
dashed lines. The cells and subcells of this realization of the dual 
complex generally are not flat where they meet a boundary between simplices.}
\label{derived_complex}
\end{figure}

The simplicial model of GR presented here makes use of a somewhat intricate 
cellular spacetime structure known to mathematicians as the  
``derived complex'' \cite{Maunder}. Spacetime is
represented fundamentally by an orientable simplicial complex $\Dg$.
\footnote{%
The simplicial complex will always be assumed to be a combinatorial 
manifold, so it has every nice property that one would expect a simplicial
representation of spacetime to have. See \cite{Schleich93} for details.}
The derived complex is defined by subdividing each 4-simplex of $\Dg$
into 10 ``corner cells'', each associated with a vertex of the simplex,
as follows. A 4-simplex $\n$ has an affine structure 
(i.e. it is a chunk of a vector space) so there is a unique constant metric 
which makes it a unit, equilateral 4-simplex. Using this metric 
the corner cell $c_P$ of the vertex $P$ in $\n$ can be defined as the 
closure of the set of points in $\n$ that are closer to $P$ than to any 
other vertex of $\n$ (See Fig. \ref{derived_complex} a)).

$c_P$ has one vertex in the interior of $\n$, namely at the center
$C_\n$ of $\n$, which is equidistant from all the vertices of $\n$.
\footnote{%
An equivalent definition of the center $C_\m$
of a simplex $\m$ is that it is the average of all the vertices of $\m$ in 
any linear coordinates on $\m$.} 
The other vertices of $c_P$ live on the 
boundary of $\n$ and thus in some subsimplex. Each subsimplex is 
equilateral, so the intersection of $c_P$ with a subsimplex $\mu$ is just
the corner cell of $P$ in $\m$ and the vertex of $c_P$ in $\m$ is the center
of $\m$. It is not hard to see that $c_P$ is topologically a hypercube and 
its vertices are $P$, $C_\n$, and the centers of all the subsimplices of $\n$
incident on $P$. (4 1-simplices, 6 2-simplices, and 4 3-simplices). 

Notice that to each vertex $P$ of a simplicial complex one can associate a 
``dual'' cell $P^*$ formed by the union of all corner cells of $P$ in 
the simplices incident on $P$. The complex, $\Dg^*$, of these dual cells is
topologically dual to $\Dg$. See Fig \ref{derived_complex} b).

\begin{figure}
\centerline{\psfig{figure=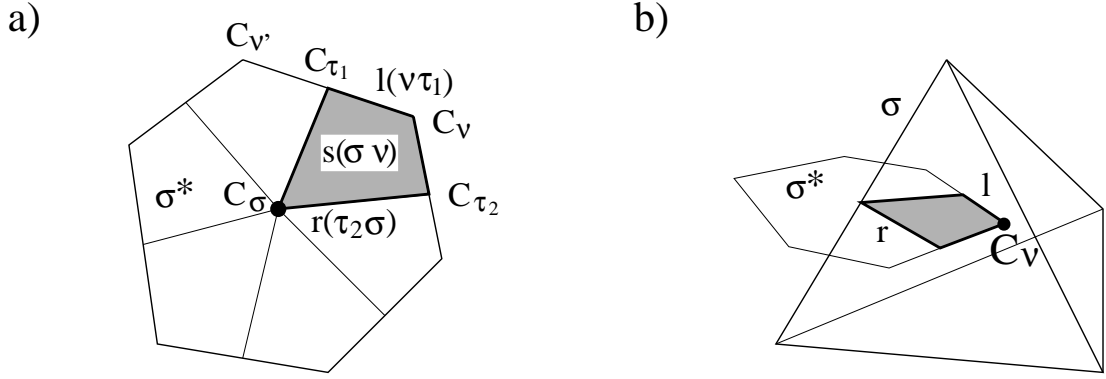,height=5cm}}
\caption[xxx]{Panel a) illustrates the definitions of $s(\sg\n)$ and the edges
$l(\n\tau)$ and $r(\tau\sg)$. In the middle lies the center of a 2-simplex
$\sg$. The corners are the centers of all the 4-simplices $\n, \n', ...$
that share $\sg$. The curve $C_\n C_{\tau_1} C_{\n'}$ connecting $C_\n$
and $C_{\n'}$ is the edge $\tau_1^*$ in $\Dg^*$ which is dual to $\tau_1$. Similarly $\cup_\n s(\sg\n)$ is the 2-cell $\sg^* < \Dg^*$ dual to 
$\sg$. One may think of
$s(\sg\n)$ as the wedge of $\sg^*$ in $\n$: $s(\sg\n) = \sg^* \cap \n$. 
Likewise, $l(\n\tau) = \tau^*\cap \n$ and the radial edge $r(\tau\sg) = 
\sg^*\cap\tau$.
\newline\newline
Panel b) shows the analogous structure in a 3-dimensional 
complex with a 3-simplex playing the role of $\n$, 2-simplices as $\tau_1$
and $\tau_2$, and a 1-simplex as $\sg$.} 
\label{cells}
\end{figure}

In the definition of the simplicial model a central role will be played by 
the $2$-cells $s(\sg\n)$ of the derived complex which are attached to the 
centers of the 4-simplices. Each is associated with a 4-simplex $\n$ and a 
2-simplex $\sg$ of $\n$, and is a plane quadrilateral formed by
the centers of $\n$, $\sg$, and the two 3-simplices $\tau_1$ and $\tau_2$ 
of $\n$ that share $\sg$. $s(\sg\n)$ can also be thought of as the 
restriction to $\n$ of the cell $\sg^*$ of the dual complex dual to $\sg$. 
See Fig. \ref{cells}.

$s(\sg\n)$ will be called a ``wedge'' and will often be denoted by just $s$
or $\sg\n$. 4-simplices will be denoted by $\n$, with some additional
subscripts or markings to distinguish different 4-simplices. 3-simplices
will similarly be denoted by $\tau$, and 2-simplices by $\sg$. 0-simplices,
i.e. vertices, will be denoted by latin capitals $P, Q, R, ...$.
Finally, $\m < \rho$ signifies that $\m$ is a subcell or subsimplex of 
$\rho$ which may be a simplex, a cell of the derived complex, a cell of the 
dual complex, or a complex.

The 4-simplices will be given a uniform orientation throughout $\Dg$, and
the orientation of each wedge $s(\sg\n)$ will be determined by the 
orientations of $\sg$ and $\n$ through the requirement that a positively
oriented basis on $\sg$ concatenated with a positively oriented basis
on $s$ forms a positively oriented basis on $\n$.

In the left-handed simplicial model of GR presented here one 
associates to each wedge $s(\sg\n)$ an $SU(2)$ 
spin 1 vector $e_{s\,i}$, which will more or less play the role of Plebanski's
$\Sg_i$ field.\footnote{%
$e_{\sg\n}$ is defined to reverse sign when the orientation of $\sg$ is 
reversed.}
The role of $A$ is played by $SU(2)$ parallel propagators along the edges of the
$s(\sg\n)$. Specifically, there is an $h_l \in SU(2)$ for each edge
$l(\n\tau)$ from the center of 4-simplex $\n$ to that of 3-simplex 
$\tau < \n$, and there is a $k_r \in SU(2)$ for each edge $r(\tau\sg)$
from the center of 3-simplex $\tau$ to that of 2-simplex $\sg < \tau$. 

Finally, a spin 2 $SU(2)$ tensor $\varphi_\n$ (represented by a symmetric, 
traceless matrix, $\varphi^{ij}_\n$) is 
associated with each 4-simplex. $\varphi^{ij}_\n$ plays the role of
$\phi^{ij}$.  

The action for the model is
\be \label{simplicial_action}
I_{\Dg} = \sum_{\n<\Dg} [ \sum_{s<\n} e_{s\,i}\theta^i_s - \frac{1}{60}
	\varphi^{ij}_\n \sum_{s,\bar{s}<\n} e_{s\,i}e_{\bar{s}\,j}
sgn(s,\bar{s})].
\ee
$\theta^i_s$ is a measure of the curvature on $s$. It is a function of the 
$SU(2)$ parallel propagators via
\be
\theta^i_s = tr[J^i g_{\di s}],
\ee
where $g_{\di s}$ the holonomy around $\di s$, and the $J_i$ are $1/2$ the
Pauli sigma matrices\footnote{%
\be
\begin{array}{ccc}
J_1 = \frac{1}{2}\left[ \begin{array}{cc} 0 & 1 \\ 1 & 0 \end{array}\right] &
J_2 = \frac{1}{2}\left[ \begin{array}{cc} 0 & -i \\ i & 0 \end{array}\right] &
J_3 = \frac{1}{2}\left[ \begin{array}{cc} 1 & 0 \\ 0 & -1 \end{array}\right].
\end{array}
\ee}
$g_{\di s}$ and $\theta_s$ may be written in terms of a rotation vector $\rho^i$ as
\bearr
g_{\di s} & = & e^{i\rho\cdot J} = \cos\frac{|\rho|}{2}\One 
+ 2i\sin\frac{|\rho|}{2} \hat{\rho}\cdot J	\\
\theta_s & = & 2 \sin\frac{|\rho|}{2}\hat{\rho}
\eearr
($\hat{\rho} = \rho/|\rho|$).
The rotation vector is essentially the curvature on the wedge $s$, and, when
the holonomy is close to one, i.e. the curvature is small, $\theta_s$
approximates the rotation vector.\footnote{%
Note that $\theta_s$ reverses sign when the orientation of $s$ reverses because the direction of the boundary $\di s$ 
reverses, which, in turn, means that $g_{\di s} \rightarrow g_{\di s}^{-1}$,
$\rho_s \rightarrow -\rho_s$, and, finally, $\theta_s \rightarrow -\theta_s$.}

$sgn(s, \bar{s}) \equiv sgn(\sg, \bar{\sg})$ is essentially the sign of the 
oriented 4-volume spanned by
the 2-simplices $\sg$ and $\bar{\sg}$ associated with $s$ and $\bar{s}$:
If $\sg$ and $\bar{\sg}$ share only one vertex (the minimum number when both 
belong to the same 4-simplex), then the orientations of $\sg$ and $\bar{\sg}$
define an orientation for $\n$, namely the orientation of the basis produced
by concatenating positively oriented bases of $\sg$ and $\bar{\sg}$.
If this orientation matches that already
chosen for $\n$ then $sgn(s,\bar{s}) = 1$. If it is the opposite
$sgn(s,\bar{s}) = -1$. If $\sg$ and $\bar{\sg}$ share 2 or 3 vertices they
lie in the same 3-plane and span no 4-volume. In this case 
$sgn(s, \bar{s}) = 0$.
	
A nice, very explicit, formula can be given for the sum in the second term
(\ref{simplicial_action}). If the vertices of the 4-simplex are numbered
1, 2, 3, 4, 5, so that 12, 13, 14, 15 form a positively oriented basis then
\be	\label{e_wedge_e}
\sum_{s,\bar{s}<\n} e_{s\,i}e_{\bar{s}\,j} sgn(s,\bar{s}) = 
\frac{1}{4} \sum_{P,Q,R,S,T \in \{1,2,3,4,5\}} e_{PQR\,i}e_{PST\,j}
\eg^{PQRST},
\ee
where $e_{PQR} = e_{s(PQR,\n)}$, and $PQR$ indicates the 2-simplex
with positively ordered vertices $P$, $Q$, $R$.

\section{Field equations and boundary terms} \label{fieldeq}

Extremization of (\ref{simplicial_action}) with respect to $h_{l(\n,\tau)}$
is most easily carried out by parametrizing variations of $h_l$ via
$h_l + \dg h_l = h_l exp[i \ag_l\cdot J]$. Then
\be	\label{alpha_deriv}
i tr[ h_l J_i \frac{\di I_{\Dg}}{\di h_l}] 
= \frac{\di I_{\Dg}}{\di \ag_l^i}|_{\ag_l = 0}.
\ee 

\begin{figure}
\centerline{\psfig{figure=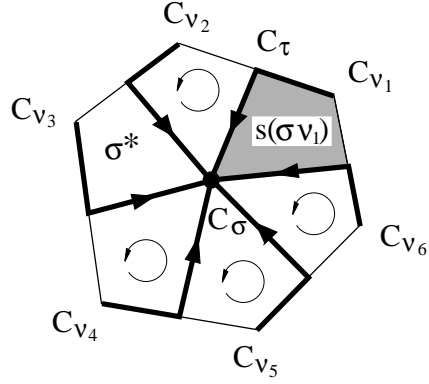,height=5cm}}
\caption[xxx]{The dual 2-cell $\sg^* < \Dg^*$ is shown with its uniform 
orientation indicated by an arrow circling in a positive sense. The routes
by which the $w_{\sg\n_n}$ are parallel transported from $C_{\n_n}$ to
$C_\sg$ to form the $u_{\sg\n_n}$ are indicated by bold lines.}
\label{transport_paths}
\end{figure}

On the other hand $l(\n,\tau) \subset \di s(\sg,\n)$ when $\sg<\tau$, 
so such a variation of $h_l$ induces
\be	\label{holonomy_var}
g_{\di \sg\n} \rightarrow g_{\di \sg\n}e^{i \ag_l\cdot J}.
\ee
(Here each $\sg<\tau$ has been oriented to match $\di\tau$ and $\tau$ 
to match $\di\n$, with the effect that $l$ is positively oriented in
$\di s(\sg,\n)$). Thus
\be	\label{theta_var}
\dg \theta^i = \ag_l^j 2 tr[J_j J^i g_{\di s}]
\ee
and
\be
\dg I_\Dg = ag_l \cdot \sum_{\sg<\tau} w_{\sg\n},
\ee
with
\be	\label{defw}	
w_{s\,j} = 2 tr[J_j J^i g_{\di s}] e_{s\,i} = 
\cos\frac{|\rho|}{2} e_{s\,j} + 2\sin\frac{|\rho|}{2}[\hat{\rho}\times e_s]_j
\ee
Extremization with respect to $h_{l(\n\tau)}$ thus requires\footnote{
If $\n = PQRST$, $\tau = PQRS$ then $\sum_{\sg<\tau} w_{s(\sg\n)} =
-w_{PQR} + w_{QRS} - w_{RSP} + w_{SPQ}$.}
\stareqn
\be	\label{h_fieldeq}
0 = i tr[ h_l J_i \frac{\di I_{\Dg}}{\di h_l}] = \sum_{\sg<\tau} w_{\sg\n\,i}.
\ee
\resteqn
(The field equations are numbered with $*$s).
Similarly, extremization with respect to $k_{r(\tau\sg)}$ requires
\stareqn
\be	\label{u_fieldeq}
u_{\sg\n_1} = u_{\sg\n_2}
\ee
\resteqn
where $\n_1$ and $\n_2$ are the two 4-simplices sharing $\tau$, and
\be 	\label{u_def}
u_{\sg\n_1\,i} = U^{(1)}[k_{r(\tau\sg)} h_{l(\n_1\tau)}]_i{}^j w_{\sg\n_1\,j}.
\ee
($U^{(1)}(g)$ is the spin 1 representation of $g\in SU(2)$).
In general $u_{\sg\n}$ is $w_{\sg\n}$ parallel transported from $C_\n$
along the boundary $\di s(\sg\n)$ in a positive sense\footnote{%
$\n_1$, $\n_2$ are numbered with index increasing in a positive sense
around $\di\sg^*$. Note that the definition of the orientation of $s(\sg,\n)$
in terms of that of $\sg$ and $\n$ defines a uniform orientation on $\sg^*$,
since the orientations of the 4-simplices is uniform in the complex $\Dg$.}
 to $C_\sg$. (See Fig. \ref{transport_paths}).
$u_s$ is thus, like $w_s$, $e$ multiplied by a factor which goes to
one as the group elements $h_l$ and $k_r$ approach $\One$. Note that
(\ref{u_fieldeq}) implies that all $u_{\sg\n}$ for a given 2-simplex
$\sg$ have a common value $u_\sg$.

Extremization with respect to $\varphi_\n^{ij}$ yields
\stareqn
\be	\label{phi_fieldeq}
\frac{\di I_\Dg}{\di \varphi^{ij}_\n} = \frac{1}{60} 
\sum_{s,\bar{s}<\n} e_{s\,i} e_{\bar{s}\,j} sgn(s,\bar{s}) \propto \dg_{ij}
\ee
\resteqn
(since $\varphi^{ij}_\n$ is traceless).

Finally, extremization with respect to $\bar{e}_s$ implies
\stareqn
\be	\label{e_fieldeq}
\frac{\di I_\Dg}{\di e_{s\,i}} = \theta^i_s - \frac{1}{30}\varphi^{ij}_\n
	\sum_{\bar{s}<\n} e_{\bar{s}\,j} sgn(s,\bar{s}) = 0.
\ee 
\resteqn

What about boundary terms? Suppose the simplicial complex has a boundary
$\di\Dg$ which doesn't cut through any 4-simplex, so it is itself a 
3 dimensional simplicial complex. No boundary term needs to be added to 
the action (\ref{simplicial_action}) if the connection is held fixed
at that boundary, that is to say, if the group elements $k_r$ on
the edges $r(\tau,\sg)$ in the boundary are held fixed. ($r(\tau\sg)$ in the
boundary is the intersection of $\tau < \di\Dg$ and the 1-cell in the dual, 
$[\di\Dg]^*$, of the the boundary which is dual to $\sg$).

Another natural field to hold fixed is $u_\sg$ for $\sg<\di\Dg$ 
(since it lives on the boundary, unlike 
$e_{\sg\n}$ and $w_{\sg\n}$ which live in the internal space
at $C_\n$, and thus always off the boundary). This corresponds more closely
to what is usually done in Regge calculus \cite{Regge61}, that is, 
holding the lengths of
the edges fixed on the boundary, because $e_{\sg\n}$, and therefore
$u_\sg$, is essentially the metrical field variable in the present simplicial 
model.
(See Appendix A for more on the relation of $e_{\sg\n}$ and $u_\sg$
to metric geometry). In this case the action must be modified: In the 
modified action the $\theta_s$ for $s$ abutting the boundary are evaluated
with $k_r$ replaced by $\One$ $\forall r<\di\Dg$ (but the definition
of $u_\sg$, (\ref{u_def}), is unchanged).

\section{The continuum limit} \label{continuum}

The form of the action (\ref{simplicial_action}) is clearly analogous
to that of the Plebanski action (\ref{Plebanski_action}). Moreover, 
Plebanski's field equations
\bearr
\frac{\dg I_P}{\dg \phi^{ij}} & = & \frac{1}{2} \Sg_i\wedge\Sg_j 
				\propto \dg_{ij}	\label{phi_Pfieldeq} \\
\frac{\dg I_P}{\dg A^i} & = & D\wedge \Sg_i = 0		\label{A_Pfieldeq} \\
\frac{\dg I_P}{\dg \Sg_i} & = & F^i - \phi^{ij}\Sg_j = 0 \label{Sig_Pfieldeq} 
\eearr
resemble simplicial field equations (\ref{phi_fieldeq}), (\ref{h_fieldeq}),
and (\ref{e_fieldeq}) respectively. We shall see that in the continuum
limit these resemblances become exact. (The simplicial field equation 
(\ref{u_fieldeq}) has no continuum analog, it is an identity in the
continuum limit I will define).

In order to take the continuum limit of the simplicial theory I define
below a map $\Omega_\Dg$ of continuum fields on a compact spacetime manifold 
$M$ into simplicial fields on a simplicial decomposition $\Dg$ of $M$, which 
allows us to represent continuum field histories by simplicial ones,
and a class of sequences $\{\Dg_n\}_{n=0}^\infty$ of simplicial decompositions
of spacetime which become infinitely fine everywhere in a nice way as
$n \rightarrow \infty$.
Any continuum field history $(A, \Sg, \phi)$ then defines a sequence of
increasingly faithful images $\Omega_n(A,\Sg,\phi) = (h,k,e,\varphi)_n$ on the complexes 
$\Dg_n$, and corresponding evaluations $I_n(A,\Sg,\phi)$ of the simplicial 
action.

I will confine myself to showing that a continuum limit of the simplicial
model reproduces the Plebanski theory for continous $\phi$ and continously
differentiable $A$ and $\Sg$. On such fields the Plebanski lagrangian,
and the functional derivatives of the action, are continous. Unless otherwise
stated $A$, $\Sg$, and $\phi$ will be assumed to satisfy these 
continuity/differentiability requirements from now on.

I will begin by showing that the evaluations $I_n$ of the simplicial action
converge to the Plebanski action in the limit of infinitely fine simplicial 
decompositions: $I_n(A,\Sg,\phi) \rightarrow I_P(A,\Sg,\phi)$ as 
$n \rightarrow \infty$. Unless $I_n(A,\Sg,\phi)$ has a more and more 
oscillatory dependence on $A$, $\Sg$, and $\phi$ as $n \rightarrow \infty$
it follows that the variations of the simplicial action due to
variations $\dg A$, $\dg\Sg$, and $\dg\phi$ of the continuum fields converge
to the corresponding variantions of the Plebanski action.
That this is in fact the case will be verified directly. The Plebanski
theory is thus the continuum limit of the simplicial theory in the weak sense
that the solutions of the Plebanski theory are the continuum field 
configurations that extremize the simplicial action
with respect to variations of the {\em continuum} fields. In terms of field 
equations, or more precisely in terms of the variational derivatives
of the actions  - which will be called the ``Euler-Lagrange functions''
(E-L functions), one can say that $A$, $\Sg$, and $\phi$ solve the
Plebanski field equations iff the simplicial E-L functions evaluated
on their image $(h,k,e,\varphi)_n = \Omega_n(A,\Sg,\phi)$ {\em integrated
against the image $(\dg h,\dg k,\dg e,\dg \varphi)_n$ of the variations
$\dg A$, $\dg\Sg$, and $\dg\phi$} vanish as $n \rightarrow \infty$.

In fact a stronger result will be proved: the E-L functions vanish more
rapidly as $\n \rightarrow \infty$ on solutions to the Plebanski equations
than on non-solutions. The simplicial E-L functions (\ref{phi_fieldeq}), 
(\ref{h_fieldeq}), and
(\ref{e_fieldeq}) on the image of a continuum field configuration turn out to 
be simply the integrals of the Plebanski E-L $d$-forms (\ref{phi_Pfieldeq}), 
(\ref{A_Pfieldeq}), and (\ref{Sig_Pfieldeq})
over suitable $d$-cells in each 4-simplex, modulo corrections of higher order
in the radius $r_\n$ of the 4-simplex (where the radius is determined by an
arbitrarily chosen positive definite background metric on spacetime). 
As a consequence the simplicial E-L functions of the image of {\em any} 
continuum field  configuration, whether solution or not, will vanish as
$n \rightarrow \infty$, simply because the cells over which the corresponding
Plebanski E-L forms are integrated get smaller as the complex is refined.
Specifically, the leading term in the E-L 
function (\ref{h_fieldeq}) vanishes as $r_\n^3$, that in
(\ref{phi_fieldeq}) as $r_\n^4$, and that in (\ref{e_fieldeq}) as $r_\n^2$. 
The integrals in each 4-simplex are sufficient to detect any non-zero
components of the E-L forms when the simplex is sufficiently small.
It therefore also follows that a continuum field configuration
is a solution of the Plebanski field equations iff all the E-L functions 
vanish more rapidly than the non-solution rate as $n \rightarrow \infty$.

That the simplicial field equations converge to those of the Plebanski theory 
in this stronger sense is significant. M. Miller \cite{Miller95} and 
Brewin \cite{Brewin95} have found that the analogous result does {\em not}
hold for the Regge model \cite{Regge61}! One can define a natural simplicial 
image of a 
metric on a spacetime $M$ by covering $M$ with a simplicial complex with 
geodesic edges and taking as the length of each edge its metric length in
$M$. What Miller and Brewin found is that on the
images of most solutions to Einstein's equations (including the Kerr
solution \cite{Miller95}) the Regge E-L functions do not vanish any more 
rapidly, as the 
complex is refined, than on non-solutions. On the other hand, Miller,
and Brewin\footnote{Brewin defines a whole class of discrete field equations
which approximate GR. One set of field equations in this class is a sum of
Regge equations.}, also showed that certain linear combinations of the Regge
E-L functions do vanish more rapidly than the non-solution rate. These 
combinations are essentially $\di I_{Regge}/\di g_{\m\n}$ in a cell built
out of several adjacent simplices. Their rapid vanishing guarantees that
GR is the continuum limit of the Regge theory in the weak sense that 
solutions (and only solutions) to GR extremize the Regge action with respect 
to variations of the continuum metric in the limit of an infinitely fine
simplicial complex.

What is going on? As the simplicial complex is refined, and becomes much finer
than the scale on which a given $\dg g_{\m\n}$ varies, the corresponding
variations of the edge lengths, $\dg l$, approach those due to a constant
$\dg g_{\m\n}$, which are highly correlated in spacetime. Thus extremization 
(modulo error terms which vanish as the complex is refined) with respect to 
continuum metric modes is
a much weaker requirement than extremization with respect to all edge length. 
It leaves many modes of the edge lengths free, because it allows many non-zero 
distributions of values of the E-L functions over the complex.
On the other hand, the fineness of the simplicial complex also restricts the
possible modes of the edge lengths in images of continuum metrics. On a 
suffiently fine complex the metric will be nearly constant, and the 
distribution of edge lengths, and similarly values of the E-L functions,
will be severely restricted simply by virtue of the fact that the edge lengths
are the image of a continuum metric. The question is then whether these 
restrictions, together with the field equations that come from extremizing with
respect to the continuum metric, are enough to imply that all the simplicial
E-L functions vanish (faster than the non-solution rate as the complex
is refined). In the Regge model the restrictions are not quite enough.
In the simplicial model presented here the analogous mechanism
does work. 

The result of Miller and Brewin seems a serious problem for path integral
quantizations based on the Regge model. A path integral over the edge lengths
will not have stationary phase trajectories corresponding to the full set of
solutions of Einsteins equations, and thus will not provide a quantization
of general relativity. One would have to integrate over some restricted set of
simplicial spacetimes, but to my knowledge no such restricted path integral 
has been defined. (How these considerations bear on dynamical triangulation
path integrals \cite{Weingarten82} based on the Regge action is unclear to me).
In contrast, a path integral over the fundamental variables of the simplicial 
model presented here is, at least on this count, a viable model of quantum
general relativity.

In the present work it is shown that the smooth solutions of the simplicial
theory, i.e. solutions which are images of $A$, $\Sg$, $\phi$ on complexes
much finer than the scale on which these fields vary, are in fact the
solutions of Plebansk's field equations.
What is not shown is that smooth boundary or initial data cannot lead to 
solutions which are highly crumpled, in addition to the smooth solutions,
or that solutions that are crumpled on the simplicial scale approximate
continuum solutions on larger scales.

Now to the details.
\newline\newline
\noindent {\bf Definition 1}: The map $\Omega_\Dg: (A, \Sg, \phi) \mapsto
(h, k, e, \varphi)$ of continuum fields on $M$ to simplicial fields on
the simplicial decomposition $\Dg$ of $M$ is defined by
\bearr
h_l & = & \calP e^{i\int_l A \cdot J}		\label{h_A} \\
k_r & = & \calP e^{i\int_r A \cdot J}		\label{k_A} \\
u_{\sg\n\,i} & = & \int_\sg v(C_\sg,x)_i{}^j \Sg_j(x)	\label{u_Sg}\\
\varphi^{ij}_\n & = & \phi^{ij}(C_\n)		\label{phi_phi}
\eearr
with $e_{\sg\n}$ defined from $u_\sg$ via (\ref{u_def}) and (\ref{defw}),
\footnote{The map (\ref{defw}) which defines $w_s$ in terms of $e_s$ is 
invertible except when the trace of the holonomy around $\di s$ vanishes.
However, when the connection $A$ is continous one may, by choosing a 
sufficiently fine simplicial complex make all holonomies around wedges
close to $1$.}
 
$\calP$ denotes path ordering, and $v(C_\sg, x) = 
U^{(1)}(\calP e^{i\int_{C_\sg}^x A \cdot J})$ is the parallel 
propagator of spin 1 $SU(2)$ vectors along a straight line from $x$ to $C_\sg$
(according to the affine structure of $\sg$).
\newline
\newline
This definition of $\Omega_\Dg$ is not the only one possible. Other maps
also lead to equivalence of the continuum limit of the simplicial theory 
and the Plebanski theory. For instance maps such that $h$, $k$, $e$, 
$\varphi$ converge to those of Definition 1 as the simplicial complex is 
refined are viable alternatives. However the
$\Omega_\Dg$ chosen here seems to lead to the cleanest proofs. 

Before considering the simplicial complexes to be used note that
we will be concerned with compact spacetimes or compact pieces of spacetimes,
which always admit finite simplicial decompositions\footnote{This follows
from the arguments on p. 488 of \cite{Schleich93}.}

As the sequence of progressively finer simplicial decompositions I will use 
{\em uniformly refining} sequences, which are defined as follows.
\newline
\newline
\noindent {\bf Definition 2}: $\{\Dg_n\}_{n=0}^\infty$ is a uniformly
refining sequence of simplicial decompositions of a compact manifold 
$M$ if 
\newline {}

1) $\Dg_0$ is a finite simplicial decomposition of $M$,

2) $\Dg_{n+1}$ is a finite refinement of $\Dg_n$,
\newline
\newline
\noindent and, in a fixed positive definite metric $g_0$ which is constant
on each $\n < \Dg_0$ (in linear coordinates on $\n$),
\newline {}

3) $r_n$, the maximum of the radii $r_\n$ of the 4-simplices $\n<\Dg_n$ 
approaches zero as $n \rightarrow \infty$,
and

4) $r_\n^4$ divided by the 4-volume of $\n$ is uniformly bounded for all
4-simplices $\n < \Dg_n$ as $n\rightarrow\infty$.
\newline {}

In Appendix B it is proven that conditions
3) and 4) are independent of the particular metric chosen.

The following lemma, also proven in Appendix B, will be useful
\newline
\newline
\noindent {\bf Lemma 1}: If $h$ is a continous $d$-form on a compact 
manifold $M$, $\{\Dg_n\}_{n=0}^\infty$ is a uniformly refining sequence
of simplicial decompositions of $M$, and $g_0$ is a positive definite metric
constant on each $\n<\Dg_0$, then $\forall \eg>0$ $N$ can be chosen 
sufficiently
large so that for any $d$-subcell $c$ of a 4-simplex $\n<\Dg_N$
\be	\label{int_estimate}
|\int_c h - \int_c h_{C_\n} | < \eg r_\n^d,
\ee
where $h_{C_\n}$ is the constant $d$-form (in linear coordinates on $\n$) 
which agrees with $h$ at $C_\n$, and $r_\n$ is the $g_0$ radius
of $\n$.
\newline {}

Lemma 1 can be conveniently restated in terms of the {\em characteristic
tensor} of $c$, which can be defined in linear coordinates $x^\ag$ on $\n$
by
\be	\label{characteristic_tensor}
t_c^{\ag_1...\ag_d} = \int_c dx^{\ag_1}\wedge ...\wedge dx^{\ag_d}.
\ee
$t_c$ might be called the coordinate volume tensor of $c$, it is the 
$d$-dimensional generalization of the coordinate length vector of an 
edge and the coordinate area bivector of a 2-cell. When $c$ is a d-simplex
with vertices $P_1, ..., P_{d+1}$ $t_c^{\ag_1 ... \ag_d} = (P_1 P_2)^{[\ag_1}
...(P_1 P_{d+1})^{\ag_d]}$. Using $t_c$ equation (\ref{int_estimate}) in 
Lemma 1 can be written as
\be
\int_c h = \int h(C_\n)_{\ag_1 ...\ag_2} t_c^{\ag_1 ... \ag_2} + O(\eg r_\n^d),
\ee
where $O(\eg r_\n^d)$ denotes a quantity $Q_c$ that vanishes faster than 
$r_\n^d$ as $n \rightarrow \infty$, that is, 
$max \{Q_c/r_\n^d | c<\n<\Dg_n\} \rightarrow 0$ as $n\rightarrow\infty$.

Now we are ready to prove that the simplicial action (\ref{simplicial_action})
converges to the Plebanski action in the continuum limit.
\newline
\newline
\noindent {\bf Theorem 1}: If $\{\Dg_n\}_{n=0}^\infty$ is a uniformly refining 
sequence of simplicial decompositions of a compact, orientable 4-manifold, $M$;
$A$, $\Sg$, $\phi$ are Plebanski fields on $M$ with $\Sg$ and $\phi$
continous and $A$ continously differentiable; and $I_n(A, \Sg, \phi)$ is the
evaluation of the simplicial action (\ref{simplicial_action}) on the 
simplicial fields $(h,k,e,\varphi)_n = \Omega_{\Dg_n}(A, \Sg, \phi)$ defined
on $\Dg_n$ by (\ref{h_A}) - (\ref{phi_phi}), then
\be
lim_{n\rightarrow\infty} I_n(A,\Sg,\phi) = I_P(A,\Sg,\phi).
\ee
\newline
\newline
\noindent {\em Proof}:
Choose a positive definite metric $g_0$ which is constant on each $\n<\Dg_0$.
By several 
applications of Lemma 1 one shows that $\forall\eg>0$  $N$ may be chosen
large enough so that 
\bearr
|\theta_{\sg\n}^i & -  & F(C_\n)^i_{\ag\bg} t_{s(\sg\n)}^{\ag\bg}| 
				< \eg r_\n^2  \\
|e_{\sg\n\,i} & - & \Sg(C_\n)_{\ag\bg\,i} t_\sg^{\ag\bg}|
				< \eg r_\n^2
\eearr
$\forall\sg<\n<\Dg_n$. Thus
\bearr
I_N & = & \sum_{\n<\Dg_N} [ \Sg_{i\,\ag\bg} F^i_{\cg\dg}|_{C_\n} \sum_{\sg<\n} 
t_\sg^{\ag\bg} t_{s(\sg,\n)}^{\cg\dg} \nonumber \\
& & \ \ \ \ -\frac{1}{60}\phi^{ij}\Sg_{i\,\ag\bg}\Sg_{j\,\cg\dg}|_{C_\n}
	\sum_{\sg,\bar{\sg}<\n} t_\sg^{\ag\bg} t_{\bar{\sg}}^{\cg\dg}
	sgn(\sg,\bar{\sg})	+ \Dg I_{N\n} ],
\eearr
where the error term $\Dg I_{N\n}$ is bounded by
\be
|\Dg I_{N\n}| < \eg r_\n^4 [ 10 || F || + (10 + \frac{1}{2} || \phi ||)
	(||\Sg || + \eg )] |_{C_\n}
\ee
(Here the norm $|| T ||$ of a tensor $T^{i_1...i_m}_{\ag_1...\\ag_n}$ is
$|| T || = (T^{i_1...i_m}_{\ag_1...\\ag_n}T^{j_1...j_m}_{\bg_1...\\bg_n}
\dg_{i_1 j_1}... g_0^{\ag_1\bg_1}...)^{\frac{1}{2}}$).
Since $F$, $\Sg$, and $\phi$ are continous on the compact manifold $M$ 
they are bounded, so $|\Dg I_{N\n}| < \eg r_\n^4 \kappa$
with $\kappa$ a finite constant.

The ratio of $r_\n^4$ to the 4-volume $V_\n$ of $\n$ is uniformly bounded.
Let this bound be $R$ then $r_\n^4 < R V_\n$ implying that 
$|\Dg I_{N\n}| < \eg \kappa R V_\n$. Therefore the sum of the errors is 
bounded by
\be
|\sum_{\n<\Dg_N} \Dg I_{N\n}| < \eg \kappa R V_M,
\ee
where $V_M$ is the $g_0$ volume of $M$, a finite number. 

Straightforward calculations show
\bearr
\sum_{\sg,\bar{\sg}<\n} t_\sg^{\ag\bg} t_{\bar{\sg}}^{\cg\dg} 
sgn(\sg,\bar{\sg}) & = & 30 t_\n^{\ag\bg\cg\dg} \\
\sum_{\sg<\n} t_\sg^{\ag\bg} t_{s(\sg\n)}^{\cg\dg} & = & t_\n^{\ag\bg\cg\dg}.
\eearr
Thus, putting everything together, we get
\bearr
I_N & = & [\sum_\n (\Sg_{i\,\ag\bg} F^i_{\cg\dg} 
- \frac{1}{2}\phi^{ij}\Sg_{i\,\ag\bg}\Sg{j\,\cg\dg})|_{C_\n} 
t_\n^{\ag\bg\cg\dg}]	+ \Dg I_N	\\
& \stackrel{\longrightarrow}{\scriptstyle{N\rightarrow\infty}} & \int_M \Sg_i\wedge F^i - 
\frac{1}{2} \phi^{ij}\Sg_i\wedge\Sg_j	= I_P(A,\Sg,\phi).
\eearr
\QED

To prove the equivalence of the continuum limit of the simplicial model
with Plebanski's theory at the level of field equations I show that
the Euler-Lagrange (E-L) functions, i.e. the variational derivatives of 
the action
with respect to the fields, are just the integrals of the Euler-Lagrange
$d$-forms of the Plebanski theory over certain $d$-cells, modulo corrections
which become negligible as the simplicial complex is refined. Then, if a
``continuum limit solution'' of the simplicial model is defined to be a
continuum field configuration, $(A, \Sg, \phi)$ such that its simplicial
images $\Omega_n(A, \Sg, \phi) = (h, k, e, \varphi)_n$ have E-L functions
that vanish faster than the volumes of the corresponding $d$-cells as 
$n \rightarrow \infty$, the continuum limit solutions are precisely the
solutions to the Plebanski field equations.

The precise statement I will prove is
\newline
\newline
\noindent {\bf Theorem 2}: Under the hypothesies of Theorem 1 and the
additional condition that $\Sg$ is continously differentiable the following 
hold for any simplicial field history corresponding to a
continuum field history via (\ref{h_A}) - (\ref{phi_phi}):   
\newline
\newline
\noindent 1) The simplicial field equation (\ref{u_fieldeq}) 
$tr[J_i k_r \frac{\di I_n}{\di k_r}] = 0$ holds identically for all $n$.
\newline
\newline
\noindent 2) $\forall \n<\Dg_n,\ \ s, l < \n$
\bearr
\frac{\di I_n}{\di \varphi^{ij}_\n} & = & \int_\n \frac{\dg I_P}{\dg \phi^{ij}}
+ O(\eg r_\n^4)	 		\label{geomet_simp_Pl}\\
\frac{\di I_n}{\di e_{s\,i}} & = & \int_s \frac{\dg I_P}{\dg \Sg_i}
+ O(\eg r_\n^2)			\label{curvature_simp_Pl}\\
i tr[h_l J_i \frac{\di I_n}{\di h_l}] & = & \int_\tau \frac{\dg I_P}
{\dg A^i} + O(\eg r_\n^3)		\label{torsion_simp_Pl}
\eearr
where $\tau$ is the 3-simplex in $\n$ dual to $l$, and in the integrals
the integrands are parallel transported to $C_\n$ along straight lines.
\newline
\newline
\noindent 3) The Plebanski field equations are fully represented by the simplicial
field equations in the sense that if the Plebanski E-L forms are not almost
everywhere zero, then on a sufficiently fine simplicial complex the
simplicial E-L functions will not all vanish.
\newline
\newline
\noindent {\em Proof}: 1) follows immediately from the definition 
(\ref{u_Sg}) of $u_{\sg\n}(A,\Sg)$.
(\ref{geomet_simp_Pl}) and (\ref{torsion_simp_Pl}) from the corresponding 
Plebanski field equations (\ref{phi_Pfieldeq}) and (\ref{A_Pfieldeq}) and
\bearr
\frac{\di I_n}{\di \phi^{ij}_\n} & = & \frac{1}{60} \sum_{\sg,\bar{\sg}<\n}
e_{\sg\n\,i}e_{\bar{\sg}\n\,j} sgn(\sg,\bar{\sg})	\\
	& = & (\Sg_{i\,\ag\bg}\Sg_{j\,\cg\dg})|_{C_\n}
	\frac{1}{60}\sum_{\sg,\bar{\sg}<\n} t_\sg^{\ag\bg} 
	t_{\bar{\sg}}^{\cg\dg}sgn(\sg,\bar{\sg}) + O(\eg r_\n^4) \\
	& = & \frac{1}{2} (\Sg_i\wedge\Sg_j)|_{C_\n\ \ag\bg\cg\dg}
		t_\n^{\ag\bg\cg\dg} + O(\eg r_\n^4)	\\
i tr[h_l J_i \frac{\di I_n}{\di h_l}] & = & \sum_{\sg<\tau} w_{\sg\n\,i}
	= h_{l\,i}^{-1}{}^j\sum_{\sg<\tau} k_{r(\tau\sg\,j}^{-1}{}^k 
							u_{\sg\,k}  \\ 
      & = & (D\wedge\Sg_i)|_{C_\n\ \ag\bg\cg}t_\tau^{\ag\bg\cg} 
						+  O(\eg r_\n^3).
\eearr
Similarly (\ref{curvature_simp_Pl}) follows from (\ref{Sig_Pfieldeq})
and the identity $\sum_{\bar{\sg}<\n} t_{\bar{\sg}}^{\ag,\bg} 
sgn(\sg,\bar{\sg}) = 30 t_{s(\sg,\n)}^{\ag\bg}$ via 
\bearr
\frac{\di I_n}{\di e_{s\,i}} & = & \theta^i_{\sg\n} 
- \frac{1}{30}\varphi^{ij}_\n
\sum_{\bar{\sg}<\n} e_{\bar{\sg}\n\,j} sgn(\sg,\bar{\sg})   \\
& = & (F^i - \phi^{ij} \Sg_j)|_{C_\n\ \ag\bg} t_{s(\sg\n)}^{\ag\bg}
+ O(\eg r_\n^2).
\eearr

3) follows as a corrollary of 2). Condition 4) in Definition 2, which
defines $\Dg_n$, prevents the 4-simplices 
from having zero volume, and therefore ensures that in each such 4-simplex
$\n < \Dg_n$ the  $t_s$, $t_\tau$, $t_\n$ span the spaces of 2, 3, and 4
index antisymmetric tensors at $C_\n$. \QED
\newline
\newline
Using (\ref{alpha_deriv}) and the fact that under variations of $A$, $\Sg$,
$\phi$ the variations of the corresponding simplicial fields $h$, $e$, 
$\varphi$ are given by
\bearr
\dg \varphi^{ij}_\n & = & \dg\phi^{ij}(C_\n) \\
\dg e_{\sg\n\,i} & = & \int_\sg \dg\Sg_i + O(\eg r_\n^2)	\\
\dg \alpha^i_l & = & \int_l \dg A^i + O(\eg r_{\n})
\eearr
(where the integrands are parallel transported to $C_\n$ along straight lines).
one finds that
\newline
\newline
\noindent{\bf Corollary}: The hypothesies of Theorem 1 and $\Sg$ continously
differentiable imply that, under variations of
$A$, $\Sg$, and $\phi$
\be
\lim_{n \rightarrow\infty} \dg I_n (A, \Sg, \phi) = \dg I_P (A, \Sg, \phi).
\ee

\section{Comments} \label{comments}

\begin{figure}
\centerline{\psfig{figure=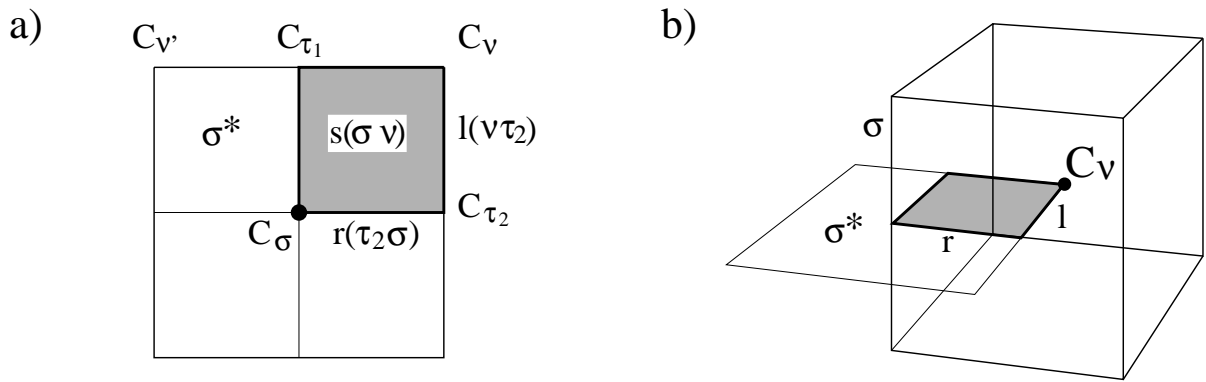,height=5cm}}
\caption[xxx]{This is the analogue of Fig. \ref{cells} for a hypercubic 
lattice. Panel a) illustrates the definitions of $s(\sg\n)$ and the edges
$l(\n\tau)$ and $r(\tau\sg)$. In the middle lies the center of a 2-cube 
$\sg$. The corners are the centers of all the 4-cubes $\n, \n', ...$
that share $\sg$. The curve $C_\n C_{\tau_1} C_{\n'}$ connecting $C_\n$
and $C_{\n'}$ is the edge $\tau_1^*$ in the lattice
$\Box^*$ dual to $\Box$, which is dual to $\tau_1$. Similarly $\cup_\n s(\sg\n)$ is the 2-cell $\sg^* < \Box^*$ dual to $\sg$. 
\newline\newline
Panel b) shows the analogous structure in a 3-dimensional 
cubic lattice with a 3-cube playing the role of $\n$, 2-cubes as $\tau_1$
and $\tau_2$, and a 1-cube as $\sg$.} 
\label{cubes}
\end{figure}

$\bullet$ A hypercubic lattice action for GR can be defined in complete
analogy with the simplicial one. One defines the centers of $n$-cubes
as the averages of their vertices, and the dual complex from these centers
as in the simplicial context. In particular, the wedges $s$, and 
the edges
$l$ and $r$ are defined by replacing $n$-simplices with $n$-cubes in their
simplicial definitions. (See Fig. \ref{cubes}).

The action of the hypercubic model is 
\be \label{hypercubic_action}
I_{\Box} = \sum_{\n<\Box} [ \sum_{s<\n} e_{s\,i}\theta^i_s - \frac{1}{8}
	\varphi^{ij}_\n \sum_{s,\bar{s}<\n} e_{s\,i}e_{\bar{s}\,j}\ 
	sgn(s,\bar{s})],
\ee
where $\Box$ is the hypercubic lattice and $\n$ labels 4-cubes. 
$sgn(s,\bar{s}) = sgn(\sg,\bar{\sg})$ is the sign of the 4-volume
spanned by the 2-cubes (squares) $\sg$ and $\bar{\sg}$ dual to 
$s$ and $\bar{s}$ respectively, provided $\sg$ and $\bar{\sg}$ share
one vertex. Otherwise $sgn(s,\bar{s}) = 0$. In particular this means that
it is zero when the 2-cubes share no vertices.

Plebanski theory is the continuum limit of the hypercubic theory 
(\ref{hypercubic_action}) in the same sense as it is the continuum limit
of the simplicial theory. (The proofs of section \ref{continuum} go through
with minor adjustments).

$\bullet$ Bostr\"{o}m, M. Miller and Smolin \cite{Bos94} found a hypercubic 
lattice
action corresponding to the CDJ action \cite{CDJ} (which is closely related
to that of Plebanski) by following the method of Regge \cite{Regge61} and
evaluating the continuum action on field configurations in which the
curvature has support only on the 2-dimensional faces of a spacetime lattice.
Unfortunately, the continuum action is not unambigously defined on such
field configurations. Nevertheless, Bostr\"{o}m {\em et. al.} present an 
action corresponding to a particular disambiguation of this expression. This 
action, written in terms of the discrete curvature variable of
Bostr\"{o}m {\em et. al.} is formally similar to that one obtains from
(\ref{hypercubic_action}) by eliminating $e_s$ using the field equations.  
However, their curvature is defined very differently from the curvature 
($\theta_s$) used here, in terms of the fundamental fields, which in their 
case is a discrete connection $1$-form and in the present case is the set of 
parallel propagators.

$\bullet$ Several authors \cite{Smolin_Renteln89}, \cite{Immirzi94}, 
\cite{Loll95}, 
\cite{Zapata96} have proposed canonical lattice formulations of general 
relativity using lattice analogs of the Ashtekar variables. It would be 
very interesting to compare the present model with these canonical
theories. This would require a canonical formulation of the present model, 
which has not yet been found. Perhaps an approach similar to that of 
\cite{Khatsymovsky95} would work.

$\bullet$ The simplicial theory presented here converges in the
continuum limit to Plebanski's theory, which 
is not equivalent to Ashtekar's canonical theory when the spatial
metric is degenerate. Thus one would expect a quantization of the present
simplicial model to approximate a quantization of the constraints of
\cite{Reisenberger95} rather than Ashtekar's constraints.

\section*{Acknowledgments}

A discussion with Jos\'{e} Zapata gave the impetus to revive this once 
abandoned
project. I am also indebted to Abhay Ashtekar, Lee Smolin, Carlo Rovelli, 
John Baez, and John Barret for essential discussions and encouragement,
and to the second referee for pointing out the work of M. Miller and Brewin.
Finally, I would like to acknowledge the support of the Center for
Gravitational Physics and Geometry and the Erwin Schr\"{o}dinger
Institute, where this work was done.

\appendix
\section{The simplicial field $e_s$ and metric geometry} \label{geometry}

On non-degenerate solutions Plebanski's fields $A$, $\Sg$, $\phi$ have
metrical interpretations. Thus simplicial fields defined via (\ref{h_A}) - 
(\ref{phi_phi}) should also have a metrical interpretation. In fact, when $\Sg_i$
satisfies the field equation (\ref{phi_Pfieldeq}) $\Sg_i\wedge\Sg_j \propto \dg_{ij}$,
and the non-degeneracy requirement $\Sg_k\wedge\Sg^k \neq 0$, it defines
a non-degenerate cotetrad $e^I{}_\ag$ (unique up to $SO(3)_L$ transformations)
and thus a non-degenerate metric\footnote{A spinorial proof of this result 
is given in \cite{CDJM}. A proof in the language of $SO(3)$ tensors is provided
in Appendix B of \cite{Reisenberger95}.}
\be	\label{metric}
g_{\ag\bg} = e^I{}_\ag \dg_{IJ} e^J{}_\bg.
\ee
Furthermore, when $A$ obeys the field equation $D\wedge \Sg_i = 0$,
the Plebanski action becomes the Einstein-Hilbert action of the metric 
(\ref{metric}),\footnote{For a proof see \cite{Reisenberger95}, Section 3.}
so this metric is the physical metric.

On solutions of (\ref{phi_Pfieldeq}) $\Sg_i = 2[e\wedge e]^{+\,0i}$, where
\be
[e\wedge e]^{+\,IJ} = \frac{1}{2} e^I\wedge e^J + \frac{1}{4}\eg^{IJ}{}_{KL}
e^K\wedge e^L
\ee
is the self-dual part of $e\wedge e$. Therefore, by (\ref{u_Sg}), the
simplicial variable $e_{\sg\n}$
is the self-dual part of the metric area bivector of $\sg$ (modulo corrections
that vanish faster than the area as the simplicial complex is refined):
\be	\label{e_is_area}
e_{\sg\n\,i} = \int_\sg \Sg_i + O(\eg r_\n^2) = 2 a_\sg^{+\,0i} +
O(\eg r_\n^2),
\ee
with $a_\sg^{IJ} = e^I{}_\ag e^J{}_\bg|_{C_\n} t_\sg^{\ag\bg}$ the metric
area bivector, equal to the coordinate area bivector in metric normal
coordinates.

If normal coordinates $x^I$ are chosen so that $\sg$ lies in a spatial
hypersurface ($x^0 = \mbox{\em constant}$) then $e_{\sg\n\,i} = \frac{1}{2}
\eg^i{}_{jk}a_\sg^{jk}$ - exactly the normal area vector of $\sg$.

It would be nice to give a metric interpretation of $e_{\sg\n}$ also away
from the continuum limit, to make possible a direct comparison of the present 
simplicial theory with Regge calculus \cite{Regge61}. 
This is difficult since,
unlike in Regge calculus, the simplices of the present model are not flat. 
The wedges, $s$, carry curvature, $\theta_s$, which does not generally vanish,
even on solutions.

However, when the holonomies $g_{\di s}$ are all $\One$
the metric interpretation of the continuum limit extends to arbitrary
simplicial complexes. 
The only non-degenerate solutions satisfying this 
requirement exactly are flat spacetimes, even though in continuum euclidean 
GR there are curved non-degenerate solutions with vanishing self-dual 
curvature. It seems that the discrete solutions approximating curved 
non-degenerate anti-self-dual solutions necessarily have some self-dual 
curvature. Nevertheless, it is interesting to see how the metric emerges
even in flat solutions.
 
With $g_{\di s} = \One$, $w_s = e_s$, so field equation
(\ref{h_fieldeq}) requires
\be	\label{div}
\sum_{\sg<\tau} e_{\sg\n\,i} = 0.
\ee
(\ref{div}) imposes 12 independent linear constraints on the 30 $e_{\sg\n\,i}$,
which imply just that there is a 2-form $\Sg_{\n\,i\,\ag\bg}$ (18 components)
such that
\be
e_{\sg\n\,i} = \Sg_{\n\,i\,\ag\bg} t_\sg^{\ag\bg}.
\ee
Field equation (\ref{phi_fieldeq}) and the non-degeneracy condition 
\be
\sum_{\sg,\bar{\sg}
<\n} e_{\sg\n}\cdot e_{\bar{\sg}\n} sgn(\sg,\bar{\sg}) \neq 0
\ee
are 
equivalent to $\Sg_{\n\,i}\wedge\Sg_{\n\,j}\propto \dg_{ij}$ with
$\Sg_{\n\,k}\wedge\Sg_\n^k \neq 0$. As for the continuum $\Sg$ fields this
implies $\Sg$ defines a non-degenerate metric, and that $e_{\sg\n}$ is the
self-dual part of the metric area bivector.

The roles played by the field equations are worth noting. A co-tetrad that 
is constant on a simplex defines an image of the 
simplex in an affine 4-space $\cal E$ with metric $\dg_{IJ}$ and a fixed 
orthonomal basis
$\{v_0,v_1,v_2,v_3\}$. (\ref{div}) ensures that any 3-simplex can be mapped
into $\cal E$ such that the of $e_{\sg\n\,i}$ of its faces are the self-dual 
parts of the area bivectors. This requirement fixes the image 
(the ``geometrical image'') of the 3-simplex up to
$SU(2)_R$ transformations. \footnote{{\em Proof}: clearly the simplex can 
be uniquely mapped
into the spatial ($123$) hyperplane of $\cal E$ such that the $e_{\sg\n\,i}$
are the spatial normal area vectors. These are equal to the self-dual parts 
of the spacetime area bivectors, which are invariant under $SU(2)_R$ 
transformations. Thus any $SU(2)_R$ transformed image of the 3-simplex
fulfills the requirement. With some more effort one can show that these are 
the only allowed transformations.}
(\ref{phi_fieldeq}) and the non-degeneracy condition ensure that the
3-simplices can all be mapped geometrically into $\cal E$ by the same
co-tetrad, so they ensure that the image 3-simplices all can fit together
to form a 4-simplex. Finally, in the gauge in which $e_{\sg\n} = u_\sg$
field equation (\ref{u_fieldeq}) ensures that the geometrical images of
the 3-simplex faces of neighboring 4-simplices match up modulo an
$SU(2)_R$ transformation. The only remaining field equation, (\ref{e_fieldeq}),
simply requires $\varphi_\n = 0$.

The field equations thus restrict the simplicial fields to ones corresponding
to a Regge type simplicial geometry determined by edge lengths. Moreover,
since the transformation needed to match up the geometrical images of the
3-simplex shared by two 4-simplices is right-handed, the holonomy around a
2-simplex of the metric compatible connection, which by definition
transports the one image of the 3-simplex into the other, is purely 
right-handed.
Since the metric compatible holonomy leaves the image of the central 2-simplex
invariant, it can only be right-handed if it is $\One$.

In a curved field configuration (\ref{h_fieldeq}) and (\ref{u_fieldeq}) contain curvature
terms which spoil the exact metrical interpretation of $e_s$ (though it
is of course recovered as the continuum limit is approached).

\section{Lemmas for the continuum limit} \label{lemmas}

\noindent{\bf Lemma B.1}: If $\{\Dg_n\}_{n=0}^\infty$ is a uniformly 
refining sequence of simplicial decompositions with repsect to one 
positive definite metric $g_0$ which is constant on each 4-simplex of $\Dg_0$,
then it is a uniformly refining sequence with respect to any other 
such metric $g_0'$.
\newline
\newline
\noindent{\em Proof}:
Since all $\Dg_n$ $n>0$ are refinements of $\Dg_0$ 
and $\Dg_0$ has a finite number of simplices it is sufficient to prove
that conditions 3) and 4) of Definition 2 hold with respect to $g_0'$ 
{\em in each 4-simplex} 
$\n_0<\Dg_0$. Inside $\n_0$ $g_0$ and $g_0'$ are constant metrics, and
since they are both positive definite they are related by a non-singular 
linear transformation. Hence there exist non-zero, finite constants
$a$ and $b$ so that 
\bearr
r'_\n & < & a r_\n\\
V_\n' & = & b V_\n.
\eearr
(${}'$ quantities are calculated with $g_0'$). 

Thus 3), $r_n \rightarrow 0$ as $n \rightarrow \infty$, implies $r_n' < a r_n$ 
also approaches zero in this limit, that is 3), also holds with respect to 
$g_0'$. Furthermore, if $\exists c>0$ such that
$V_\n > c r_\n^4$ then $V_\n' = bV_\n > cb r_\n^4 > cb/a^4 r_\n'^4$,
so 4) with respect to $g_0$ implies 4) with respect $g_0'$. \QED
\newline
\newline
\noindent{\bf Lemma A.2}: If $f$ is a continous function on a compact 
manifold $M$ and $\{\Dg_n\}_{n=0}^\infty$ is a uniformly refining sequence
of simplicial decompositions of $M$ then $\forall \eg>0$ $N$ can be chosen
sufficiently large that $|f(x) - f(y)|<\eg$ $\forall x,y\in\n<\Dg_N$.
\newline
\newline
\noindent{\em Proof}: Since the simplicial decomposition $\Dg_0$ is
finite, it is sufficient to prove the lemma in each $\n_0<\Dg_0$ separately.
The continuity of $f$ implies that, with respect to any constant, positive 
definite metric $g_0$ on $\n_0$, there exists for each point $x\in\n_0$ an 
open ball $B(x,r_x)$ of radius $r_x>0$ about $x$ such that $|f(x) - f(y)|<
\eg/2$ $\forall y\in B(x,r_x)$. 

Since $\n_0$ is compact it can be covered by a finite set of balls 
$\{B(x_m,1/3 r_{x_m})\}_{m=1}^M$. Now let $r = 1/3 max\{r_{x_m}\}$,
and note that $\forall x \in \n_0$ $B(x,r) \cap \n_0 \subset B(x_m, r_m)
\cap \n_0$ for some $m \in \{1,...,M\}$. This implies that 
$|f(x) - f(y)| < \eg$ $\forall x, y \in B(x, r)\cap \n_0$. Since
N can be chosen large enough that $r_\n < r$ $\forall \n<\Dg_N \cap \n_0$ 
the requirements of the lemma are satisfied in $\n_0$. \QED
\newline
\newline
\noindent{\bf Corrollary} (Lemma 1): If $h$ is a continous $d$-form on 
$M$ and $g_0$ is a positive definite metric adapted to $\Dg_0$, 
then $\forall \eg>0$ $N$ can be chosen sufficiently
large so that for any $d$-cell $c<\n<\Dg_N$
\be
|\int_c h - \int_c h_{C_\n} | < \eg r_\n^d,
\ee
where $h_{C_\n}$ is the constant $d$-form (according to the affine structure
of $\n$) which agrees with $h$ at $C_\n$, and $r_\n$ is the $g_0$ radius
of $\n$.
\newline
\newline
\noindent{\em Proof}: Fix on each $\n_0$ normal coordinates $x^\ag$ of 
$g_0$. 
\be	
|\int_c h - \int_c h_{C_\n} | < \sum_{\ag_1...\ag_d} \int_c |h(x)_{\ag_1...\ag_d}
-h(C_\n)_{\ag_1...\ag_d}| |d x^{\ag_1}...d x^{\ag_d}|.
\ee
Furthermore, since the components $h_{\ag_1...\ag_d}$ are continous functions,
one may choose $N$ large enough that $|h(x)_{\ag_1...\ag_d}
-h(C_\n)_{\ag_1...\ag_d}| < \eg (\mbox{\em dimension $M$})^{-d} 2^{-d}$. Thus
\bearr
|\int_c h - \int_c h_{C_\n} | & < & \eg max 
\{ \int_c |d x^{\ag_1}...d x^{\ag_d}|\}		\\
	& < & \eg r_\n^d.
\eearr
\QED


\end{document}